# Observation of Cloud-to-Ground Lightning Channels with High-Speed Video Camera

Buguet M.[1] *, P. Lalande[1], P. Blanchet[1], S. Pédeboy[2], P. Barneoud[2], P. Laroche[1]

1. ONERA – DMPH – 29 avenue de la division Leclerc, 92290 Châtillon, France
2. METEORAGE – Hélioparc, 2 avenue Pierre Angot, 64053 Pau, France

**ABSTRACT:** Between May and October 2013 (period of sustained thunderstorm activity in France), several cloud-to-ground lightning flashes have been observed in Paris area with a high-speed video camera (14000 frames per second). The localization and the polarity of the recorded cloud-to-ground flashes have been obtained from the French lightning detection network Météorage which is equipped with the same low frequency sensors used by the US NLDN. In this paper we focused on 7 events (3 positive cloud-to-ground lightning flashes and 4 negative cloud-to-ground lightning flashes). The propagation velocity of the leaders and its temporal evolution have been estimated; the evolution of branching of the negative leaders have been observed during the propagation of the channel which get connected to ground and initiate the first return stroke. One aim of this preliminary study is to emphasize the differences between the characteristics of the positive and of the negative leaders.

## INTRODUCTION

The first stroke of a downward negative cloud to ground flash (-CG) is preceded by a stepped leader. The temporal delay between each step can varies from 5 to 100 µs and the length have a typical value of 50 m (ranging from 5 to 200 m according to the studies) [Berger 1967; Chen et al. 1999; Hill et al. 2011]. The propagation velocity of the stepped leader determined from photographs ranges from 1 to $25 \times 10^5$ m s$^{-1}$ (with an averaged around $2 \times 10^5$ m s$^{-1}$) [MacGorman and Rust, 1998]. Generally, the first stroke of a downward positive cloud-to-ground lightning flash (+CG) is preceded by a positive leader which propagates continuously, with a propagation velocity ranging from $10^4$ m s$^{-1}$ and $10^6$ m s$^{-1}$ [Les Renardières Group 1977; Campos et al. 2014].

During the experiment 150 videos of CG flashes were recorded. The quality of the data depends on the luminosity of the leader channel: it may be altered by the occurrence of precipitation and by the distance between the leader and the camera.

## DATA

### The videos

We use a high-speed video camera (14000 frames per second) installed in our Lab at the ONERA center of Châtillon, South of Paris. The field of view of the camera is around 40°. We focus on 7 videos of CG flashes which occurred in July 2013: for 6 of them, the camera is oriented to the North-East and for the 7$^{th}$ (n°4) the camera is oriented to the South-East. The camera is a phantom V711 camera, which works in the visible range, the delay between two pictures is 71.43 µs with a time of exposure of 70.24 µs, thus the camera is blinded only 1.7 % of the time. The duration of the recorded videos is of 76 ms (i.e. 1068 pictures) with a resolution of 1024 × 512. The videos are synchronized with a GPS time reference.

---

* Contact information: Magalie BUGUET, ONERA, Châtillon, France, Email : magalie.buguet@onera.fr

*The lightning flashes locations*

To localize the lightning flashes recorded by the camera, the data from Météorage, the French national lightning detection network, are used. This network (member of EUCLID network) is composed of 19 sensors installed all around France. International cooperations with neighboring countries allow Météorage to extend its coverage to a large part of Western Europe. The network uses the LS7001 sensors from Vaisala, measuring in real time the angle of incidence and the time of arrival of signals generated by return strokes. The latter are then located by a position analyzer which combines both the magnetic direction finding and time of arrival techniques to find the location of return or subsequent strokes in flashes. In addition to the position and the time of occurrence, the polarity and the peak current amplitude are determined for each stroke. The sensors have a precise clock, periodically synchronized on the Global Positioning System satellite (GPS), whose accuracy is around 100 ns. The respective mean flash and stroke detection efficiencies in France are 96 % and 89 % [Pédeboy and Schulz, 2014]. The Intra-Cloud detection efficiency was estimated during the Special Observation Period campaign of the HyMeX project in South-East of France to 47 % [Pédeboy, 2012]. The relative location accuracy is estimated to be better than 150 m [Schulz et al., 2014] which is consistent with results obtained on networks using the same Vaisala's technology like on the Gaisberg tower in Austria based on ground truth data [Diendorfer et al., 2014].

**METHODOLOGY**

The association between the videos and the flash location is established based on the time of each event. The impact location which is the closest from the lightning flash recorded according to a temporal delay and an estimated geographical position is attributed to the CG on the video. By this way it is possible to know the location of the CG return stroke, its peak current and polarity and to know multiplicity of the return stroke.

To estimate the length of propagation of the leader and its velocity from the video picture, we consider the closest building (of which we know the height) from the lightning flash. Considering that the building and the lightning flash are roughly at the same distance from the ONERA, we measure on the video the size of the building and established a correspondence between the measure in centimeter and the real size in meter.

The distance between the camera and the CG flash recorded have been calculated based on the location from the lightning mapper network. Based on the optical observations of the videos, an average propagation velocity is calculated for each leader by using the total height of the channel and the delay between its first apparition on the video and the connection to the ground. We have also calculated the propagation velocity per segment for each leader by evaluating the distance and the time between two segments. The segments used and the associated parameters to evaluate the temporal evolution of the propagation velocity of the leaders are determined by the following method:
- for the positive leader : the distance covered by the leader between each picture is measured;
- for the negative leader : the distance between two ramifications created by the main branch of the leader (i.e. the part of the leader which will connect to the ground) and the delay to cover this distance is measured.

Thus, the averaged propagation velocity for the negative leaders is evaluated for the main channel (i.e. the one which connects to the ground). We can also note that the propagation velocities given in this study represent a minimum velocity because the leader is projected on a 2D-plan normal to the axe of view of the camera.

The branching of the negative leaders is also studied. To do this, the number of ramification created by the leader is counted, and the branches which disappear (on the video) are differentiated from the ones which remain visible. That is to say that we counted each time when the leader split in two or more directions.

**RESULTS**

Seven videos have been analysed from now: 3 observations of +CG lightning flashes and 4 of -CG lightning flashes. The events are listed in the Table 1. They have been selected because of the good visibility of the leaders.

Table 1: Description of the observed cloud-to-ground lightning flashes

| Index | Date | Hour (UT) | Polarity | Peak Current | Multiplicity | Latitude | Longitude |
|---|---|---|---|---|---|---|---|
| 1 | 23/07/2013 | 12:54:18. 262029442 | + | 91.51 | 1 | 48.9147 | 2.3773 |
| 2 | 23/07/2013 | 13:31:38. 957271097 | + | 92.35 | 1 | 48.887 | 2.3884 |
| 3 | 26/07/2013 | 04:01:05. 649918556 | + | 76.7 | 1 | 48.8849 | 2.3741 |
| 4 | 26/07/2013 | 22:25:22. 262757233 | - | 13.25 | 2 | 48.6982 | 2.4031 |
| 5 | 27/07/2013 | 03:59:59.285128377 | - | 7.1 | 2 | 48.8315 | 2.2906 |
| 6 | 27/07/2013 | 04:01:19. 872782468 | - | 16.61 | 2 | 48.8355 | 2.2956 |
| 7 | 27/07/2013 | 04:24:55. 873289896 | - | 9.77 | 2 | 48.8155 | 2.2909 |

*Temporal evolution of the propagation velocity of leaders*

The parameters of each leader calculated from the videos are summarized in the Table 2.

For leaders of both polarities, we observe that the propagation velocities are coherent with published studies [Campos et al. 2014; Wang and Takagi 2011]. An acceleration of the propagation of the leaders when they come closer and closer to the ground is observed. This may be due to the local increase of the electric field at the head of the leader which become more and more conductive. These accelerations can be viewed in Table 2 by focusing on the minimum and maximum propagation speed per segment indicated in the two last columns. We observe that the range of values of the propagation velocity is similar between the positive and negative leader (around $10^5$ m s$^{-1}$) except in the last part of the development of the channel where the positive leaders are faster than the negative leaders (propagation speed around $10^6$ m s$^{-1}$).

The duration of the visible continuing current which remained on certain videos is also estimated. We note continuing current after the 3 observed +CG, and after 2 of the 4 observed –CG. This phenomenon seems not linked to the polarity or to the amplitude of the peak current. Note that sometimes the video ends before the disappearance of the continuing current.

Table 2: Main characteristic of the observed leaders. Min (Max) stands for minimum (maximum) values.

| Index | Continuing current (Yes/No – duration (ms)) | Distance from Lab (km) | Number of ramification (first number is for the main channel / second number is for all channels) | Number of subsisting new branch | Averaged propagation speed (m s$^{-1}$) | Propagation speed per segment | |
|---|---|---|---|---|---|---|---|
| | | | | | | Min | Max |
| 1 | y 5.609 | 14.71 | 0 | 0 | $2.98 \times 10^5$ | $4.19 \times 10^5$ | $1.96 \times 10^6$ |
| 2 | y 48.351 | 12.56 | 0 | 0 | $2.81 \times 10^5$ | $1.44 \times 10^5$ | $2.02 \times 10^6$ |
| 3 | y 60.776 | 11.76 | 0 | 0 | $3.31 \times 10^5$ | $3.49 \times 10^5$ | $2.09 \times 10^6$ |
| 4 | y 30.246 | 14.10 | 14 / 41 | 8 | $2.13 \times 10^5$ | $1.41 \times 10^5$ | $2.59 \times 10^5$ |
| 5 | y 18.389 | 3.81 | 29 / 74 | 14 | $1.95 \times 10^5$ | $1.40 \times 10^5$ | $5.20 \times 10^5$ |
| 6 | n | 4.31 | 11 / 47 | 6 | $3.45 \times 10^5$ | $1.59 \times 10^5$ | $5.98 \times 10^5$ |
| 7 | n | 2.07 | 18 / 33 | 5 | $1.14 \times 10^5$ | $8.71 \times 10^4$ | $2.79 \times 10^5$ |

*Evolution of the creation of ramification by the negative leaders*

As we can see on the Figure 1, the most important difference observed about the propagation of the +CG and –CG leader is the creation of new branches: the negative leader (Figure 1a) is ramified, and more or less luminous (characteristic of the "stepped leader") whereas the positive leader (Figure 1b) is non-ramified and seems to follow a "single" trajectory (characteristics of a continue propagation). By focusing on the optical characteristics of the negative leaders videos we note that their propagation is visible through the head of the streamers which is very luminous compared to the branches (cf. Figure 1a, first image). Then, it seems to exist a competition between the branches until one of them becomes more luminous (and probably more conductive), which can creates a kind of shielding of the electric field preventing the other branches to continue their propagation, and finally lead to the connection to the ground. The branches which are not visible during all the propagation of the leader (except their head) are all re-illuminated by the propagation of the upward return stroke and remain visible during 2 or 3 frames after the connection to ground.

However we can note, that the +CG leader studied have sometimes shown some ramifications at the top of the screen (i.e. higher than 2-3 km of altitude); this branching is evidenced by the re-illumination of the leader channels by the return stroke. This could be due to an intra-cloud component of the CG that is not recorded by the video, or to a ramification not enough visible to appear on the video. This last hypothesis could be supported by the video n°2 of which is extracted the Figure 1b, where in the first image (T-92) two branches are visible, only one on the second image (T-44) and at least 5 once the connection to ground is established (third image, T-2).

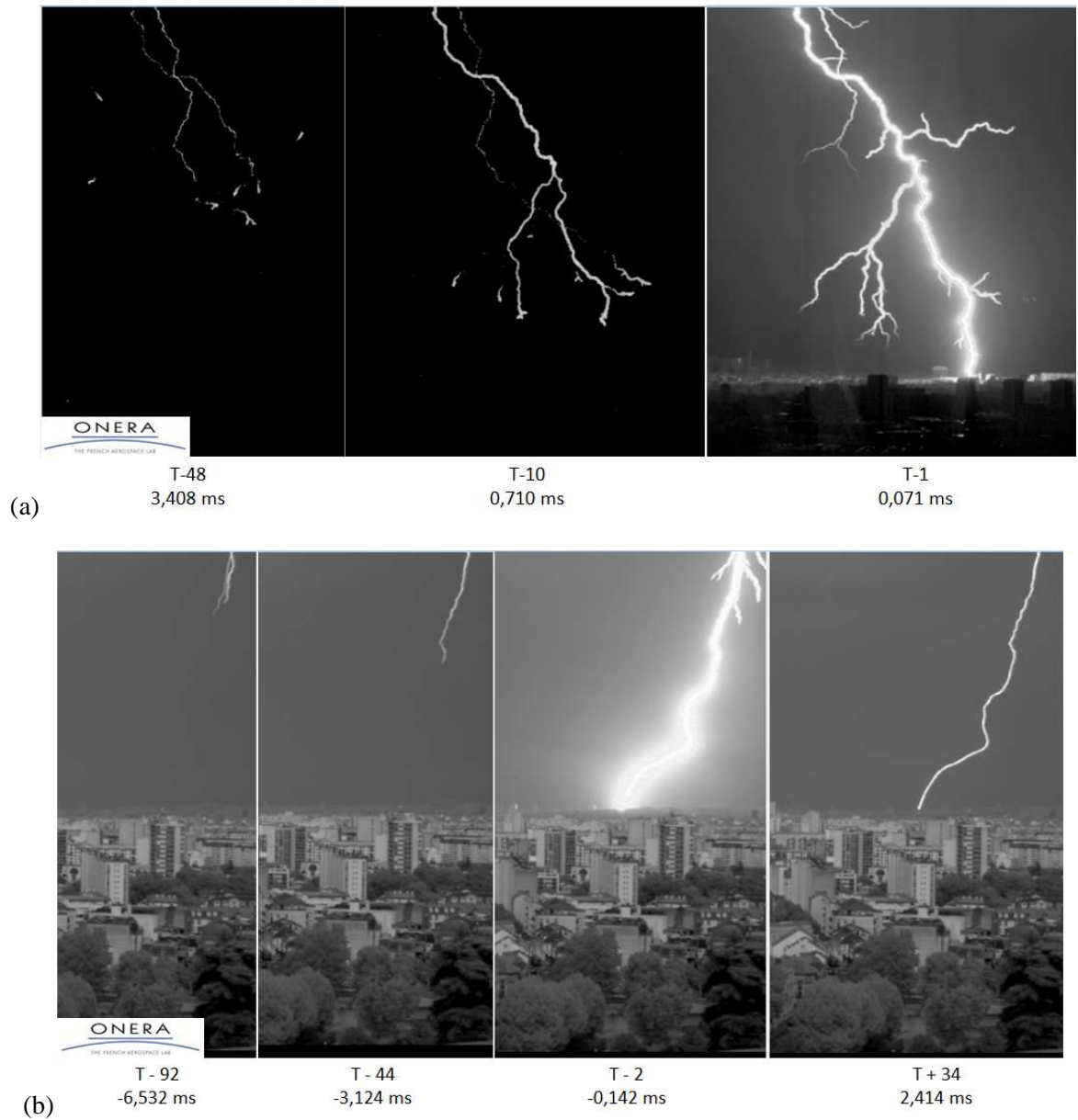

*Figure 1: Screenshots of two of the studied leaders: (a) negative cloud-to-ground lighting flash, n°5, 27/07/2013 and (b) positive cloud-to-ground lightning flash, n° 2, 23/07/2013. Under the screenshots are indicated the number of the frame before the triggered image (T), and the time of the frame is indicated in millisecond.*

Based on the 4 –CG videos studied, we observe that between 27 and 57 % of the ramifications lead to the creation of new subsisted branches connected to the main channel. Most of the time the ramification are double (the leader is separated in two new branches), but rarely and close to the ground 3 sprigs can be created with often only one or two that subsist. Nevertheless, based on the videos, we cannot determined why a branch will subsist and why another will not, as well as, it is difficult to explain what is the branch which will connect to the ground.

**CONCLUSIONS**

This preliminary study is the first step of the constitution of a video data set from which many information can be extract to better understand the propagation of the CG lightning and to better parameterized the model of lightning flash propagation. In the future we plan to investigate the evolution of the branching angular orientation of the downward negative stepped leader.